\documentclass[12pt]{article}
\usepackage{amsmath,amssymb,amsthm,amsfonts,cases}
\usepackage{parskip,url}
\usepackage[colorlinks=true,allcolors=blue]{hyperref}
\usepackage{graphicx}
\usepackage{pdflscape}
\usepackage{rotating}
\usepackage{bm}

\usepackage{authblk}

\usepackage{booktabs}
\usepackage{threeparttable} 
\usepackage[labelsep=period,font=small]{caption}
\usepackage{rotating} 
\usepackage{multirow}
\usepackage{colortbl} 
\usepackage{etoolbox}
\AtBeginEnvironment{tablenotes}{\footnotesize} 

\usepackage{epstopdf} 
\usepackage{epsfig}
\usepackage{tikz}

\usepackage{url}

\usetikzlibrary{arrows.meta, arrows,calc,patterns,shapes.geometric,automata,positioning}
\tikzstyle{state} = [circle, minimum size=1cm, text centered, draw=black]
\tikzstyle{startstop} = [rectangle, rounded corners, minimum width=3cm, minimum height=1cm,text centered, draw=black, fill=red!30]
\tikzstyle{process} = [rectangle, minimum width=3cm, minimum height=1cm, text centered, draw=black, fill=blue!30]
\tikzstyle{decision} = [diamond, minimum width=3cm, minimum height=1cm, text centered, draw=black, fill=green!30]
\tikzstyle{arrow} = [thick,->,>=stealth]

\usepackage[authormarkup=none]{changes}
\definechangesauthor[name={Mengyi}, color=orange]{MX}
\definechangesauthor[name={Ben}, color=magenta]{BL}

\DeclareMathOperator{\E}{\mathbb{E}}

\DeclareMathOperator{\Z}{\mathbf{Z}}

\DeclareMathOperator{\z}{\mathbf{z}}
\DeclareMathOperator{\sgn}{sgn}

\usepackage{natbib}

\usepackage[margin=1in]{geometry} 
\newcommand{\dif}{\mathrm{d}} 

\title{Fair Pricing in Long-Term Insurance: \\ A Unified Framework}

\author[1]{Hong Beng Lim\thanks{Corresponding author. E-mail: \url{benhblim@cuhk.edu.hk}}}
\author[2]{Mengyi Xu}
\author[3]{Kenneth Q. Zhou}

\affil[1]{\normalsize Department of Finance, Chinese University of Hong Kong, China}
\affil[2]{\normalsize School of Risk and Actuarial Studies, UNSW Sydney, Australia}
\affil[3]{\normalsize Department of Statistics and Actuarial Science, University of Waterloo, Canada}

\date{\today}

\begin{document}
\maketitle

\begin{abstract}

Extant literature on fair pricing methods for actuarial contexts has primarily focused on the regression setting. While such approaches are well-suited to short-term products, it is unclear how they generalize to long-term products, whose pricing essentially relies on estimating transition rates in multi-state models. To address this gap, we propose a unified framework that recasts the estimation of any given multi-state transition model as a set of Poisson regression problems. This reformulation enables the direct application of existing fair pricing methods, which together constitute our proposed methodology. As an illustration, we apply the framework to a fair pricing exercise for a stylized long-term care insurance product using data from the University of Michigan Health and Retirement Study (HRS), focusing on a post-processing approach. We further explain how the framework readily accommodates pre-processing and in-processing fairness methods.

\end{abstract}

\pagebreak

\section{Introduction}

Insurance pricing lies at the intersection of risk classification and fairness, raising difficult questions about how premiums should be constructed. A growing literature has sought to address this tension by proposing fair pricing methodologies, which can broadly be divided into two main strands. The first line of research focuses on the concept of the discrimination-free premium pioneered by \citet{lindholm2022discriminationfree}. Their formula is designed to mitigate the indirect impact of the discriminatory covariate on the model. Building on this work, \citet{araizaiturria2024discriminationfree} explain the formula within a causal framework, identifying scenarios where its application may be socially acceptable or inappropriate. Additional extensions of this approach have been developed for settings where information on the discriminatory covariate is incomplete \citep{lindholm2024multitask} or unavailable \citep{gabric2024bayesian}.

Another line of research draws on notions of fairness developed in the machine learning literature, as well as methods designed to enforce such notions \citep[see][for a review]{pessach2022review}. In the actuarial context, \citet{grari2022fairml} use a penalized neural network approach to impose fairness constraints, specifically addressing continuous discriminatory covariates. \citet{xin2024antidiscrimination} examine different levels of non-discrimination required by regulation and identify suitable machine learning methodologies for each level. Optimal transport methods have also been explored: \citet{lindholm2024proxy} use them to pre-process the data toward enforcing demographic parity, whereas \citet{hu2024seqfair} apply them to post-process predictions to enforce sequential fairness in multiple sensitive attributes. \citet{cote2024fair} propose a comprehensive causal framework to classify existing fair pricing methodologies in terms of the type of fairness enforced.

The aforementioned body of research has largely focused on incorporating fairness constraints into predictions of expected loss, in contrast to traditional machine learning settings where fair decision-making (i.e., classification) is of interest. As a result, methodologies have primarily been developed for the regression setting, with the loss variable serving as the response variable. However, this setup becomes limiting in the context of long-term insurance products. It is primarily suited to short-term products, such as home or automobile insurance, where premiums correspond directly to expected losses. For long-term insurance products, including life insurance and long-term care insurance (LTCI), premiums are instead derived from expected present value calculations that depend on estimates of transition rates in an underlying multi-state model, such as mortality rates in life insurance or multiple health-state transition rates in LTCI. In these settings, there is typically no single outcome variable whose expectation directly corresponds to the final price, making it unclear how existing fair pricing methods generalize beyond short-term products.

Two additional features of long-term products further underscore the importance of fairness considerations in this context. First, unlike short-term insurance products that are often purchased by companies or organizations, such as in commercial insurance and group health insurance, long-term products are typically issued to individuals. As a result, fairness concerns are more directly tied to individual policyholders in the long-term context.  Secondly, while short-term products are often renewed annually, long-term policies often span over decades. The impact of a possibly unfair premium is far more enduring in the latter case, potentially persisting throughout the policyholder’s lifetime. These features highlight the importance of addressing fairness at the outset when pricing long-term insurance products.

The distinctive features of long-term insurance pricing, together with the lasting individual consequences of unfair premiums, motivate a careful examination of how fair pricing methodologies can be applied to long-term products. This has become even more pressing in the age of big data and artificial intelligence, as the use of complex predictive models has intensified regulatory concerns about algorithmic fairness. In particular, several U.S. jurisdictions have introduced laws that require insurers utilizing external consumer data to assess whether such use has resulted in unfair outcomes for protected groups. Notable examples include Colorado's Senate Bill (SB) 21-169 \citep{colorado2022} and New York's Circular Letter No. 7 \citep{nydfs2024}. 

Given that fair pricing methods are readily applicable in the regression setting, one natural way to extend them to the long-term context is to reformulate the estimation of transition rates in multi-state models as a set of regression problems. This is precisely the approach we adopt. Specifically, we build upon the formulation dating back to \citet{sverdrup1965multipois} and further developed in \citet{renshaw1995multipois,wang2022multinn}, that the number of transitions for a given group over a prespecified time interval follows a Poisson distribution. We lend further support to this formulation using the Poisson-survival connection, and use it to construct a unified framework in which any multi-state model can be expressed as a series of Poisson regression models. Building on this framework, we propose a fair pricing procedure for long-term insurance products by first expressing the estimation of the underlying multi-state model in terms of its constituent Poisson regressions, and then applying a chosen fair pricing method to each regression component. 

Our contributions are threefold. First, we make a methodological contribution by showing that a general multi-state modeling problem can be re-expressed as a set of Poisson regression models, with transition counts treated as outcomes and exposure times as offsets. This representation provides a unified and transparent statistical structure for estimating transition probabilities in long-term insurance settings. Moreover, this reformulation allows practitioners working on long-term pricing problems to leverage well-developed Poisson regression tools that are more commonly used in short-term insurance contexts.

Second, we make a conceptual contribution by proposing a framework that enables fair pricing methods developed for short-term insurance products to be applied in long-term settings. Specifically, we describe how representative approaches from each of three commonly used fairness adjustment categories---pre-processing, in-processing, and post-processing---can be adapted to the long-term pricing context in a coherent manner. Utilizing our unified representation, the proposed framework accommodates fairness adjustments without altering the underlying multi-state structure. 

Third, we make a practical contribution by demonstrating how the proposed framework can be implemented through an LTCI pricing exercise using data from the University of Michigan Health and Retirement Study (HRS). Our case study illustrates how fairness adjustments applied at the transition level can be incorporated within a multi-state model and reflected in the resulting premiums of LTCI products. This exercise serves as a concrete illustration of the mechanics of fairness adjustments in long-term insurance pricing.

The rest of this paper is organized as follows. Section~\ref{sec:longterm} reviews a general pricing approach for long-term insurance products. Section~\ref{sec:fairpricing} summarizes existing fair pricing methods and discusses the challenges of adapting them to long-term contexts. Section~\ref{sec:unified} introduces a unified framework that presents a pathway toward overcoming these challenges. The proposed fair pricing framework for long-term insurance is then illustrated through a case study in Section~\ref{sec:casestudy} and presented in a more general form in Section~\ref{sec:longtermfair}. Finally, Section~\ref{sec:discussion} concludes with a discussion of the broader implications of the proposed framework and directions for future research.

\section{Pricing for Long-Term Insurance Products}
\label{sec:longterm}

A variety of long-term insurance products is available on the market, including life insurance, disability insurance, critical illness insurance, and LTCI. These products share several common features. First, benefit payouts are contingent upon the insured either transitioning to, or remaining in, a particular health state. Second, premiums collected in earlier periods are typically invested to fund coverage in later periods, so that pricing relies on expected present value calculations. Third, the duration of the policy is central to its value proposition, either because benefits are paid as a stream over time or because substantial future costs are spread across earlier periods.

The expected present value calculation involves transition probabilities between the health states of interest. We first describe how these multi-state transition models are set up and estimated, and then explain how the transition rates are used to price these products.

\subsection{Multi-state transition models } \label{sec:multistate}

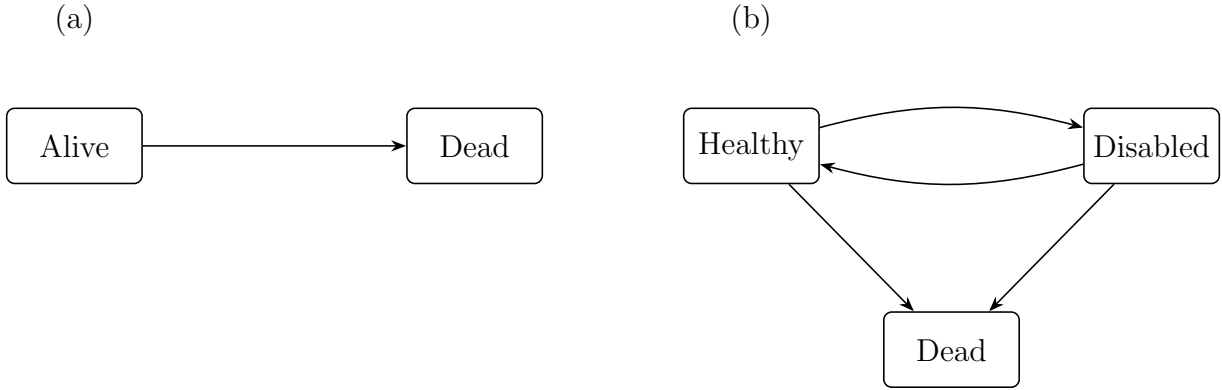
\begin{figure}[htbp]
    \centering
\begin{tikzpicture}[
  >=Stealth,
  semithick,
  node distance=25mm and 35mm,
  state/.style={
    rectangle,
    draw,
    rounded corners=3pt,
    minimum width=18mm,
    minimum height=10mm,
    inner sep=2pt
  },
]

\begin{scope}
  \node[state] (alive) {Alive};
  \node[state, right=of alive] (dead) {Dead};
  \draw[->] (alive) -- (dead);

  \node[above=8mm of alive] {(a)};
\end{scope}

\begin{scope}[xshift=90mm]
  \node[state] (H) {Healthy};
  \node[state, right=of H] (F) {Disabled};
  \node[state, below=22mm of $(H)!0.5!(F)$] (D) {Dead};

  \draw[->, bend left=15]  (H) to (F);
  \draw[->, bend left=15]  (F) to (H);
  \draw[->] (H) -- (D);
  \draw[->] (F) -- (D);

  \node[above=8mm of H] {(b)};
\end{scope}

\end{tikzpicture}
\caption{Multi-state transition models. \textit{Note}: Panel (a) depicts the two-state model for pricing life insurance and annuity products. Panel (b) depicts a three-state model with recovery from disability.}
\label{fig:multistateModel}
\end{figure}

Figure~\ref{fig:multistateModel} illustrates two multi-state models commonly used in the pricing of long-term insurance products. One of the most commonplace of such products, life insurance, may be viewed as being priced using a two-state model: this is shown in the left panel. There is a single possible transition type, as recovery from death is impossible. More generally, multi-state models are set up with three or more states, where returning from a state may be permitted. The right panel depicts such a model. This is a three-state model with healthy, disabled, and dead states, in which recovery from disability is permitted, resulting in four possible transition types. Depending on the application, alternative model specifications may restrict recovery or incorporate additional health states. 

Transitions between states are usually characterized by their transition intensities, which are modeled as functions of risk factors. A common specification assumes that each transition intensity follows a log-linear form in covariates:
\begin{equation} \label{eq:intensity}
\ln \lambda_{k, m}(t) =
	\bm{\beta}^{\prime}_m \cdot \mathbf{z}_k +
	\boldsymbol{\gamma}^\prime_m \cdot \boldsymbol{\omega}_k(t),
\end{equation}
where, for the $k$\textsuperscript{th} individual, $\mathbf{z}_k$ denotes the vector of static covariates, and $\bm{\omega}_k(t)$ denotes the vector of time-varying covariates at time $t$. The model parameters, $\bm{\theta} = (\bm{\beta}, \boldsymbol{\gamma})^\prime$, are then estimated using the maximum likelihood estimation (MLE) method.

To this end, we formulate the log-likelihood function for the multi-state model. Suppose there are $ K $ individuals, $ M $ transition types, and $ I $ observation occasions. Let $ t_{k,i} $ denote the time of the $i^\text{th}$ observation for the $k^\text{th}$ individual, and let $ \hat{t}_{k,i} $ denote the time at which a transition occurs between the $i^\text{th}$ and the $(i+1)^\text{th}$ observations, if any. Define the indicator variable $ T_{k, m, i} $, which equals 1 if a transition of type $ m $ occurs in the interval between $t_{i}$ and $t_{i+1}$, and 0 otherwise. In addition, let $ R_{k, m}(t) $ indicate whether the $k^\text{th}$ individual is exposed to the risk of experiencing transition type $ m $ at time $ t $. 

Based on these definitions, the log-likelihood function for the multi-state model can be written as
\begin{equation}\label{eq:likelihood}
	l \left(\boldsymbol{\theta} \right)= 
	\sum\limits_{k=1}^{K}
	\sum\limits_{m=1}^{M} 
	\sum\limits_{i=1}^{I-1}
	  l_{k,m,i}\left(\boldsymbol{\theta} \right),
\end{equation}	
where the contribution of individual $k$, transition type $m$, and observation occasion $i$ is given by
\begin{equation*} 
\begin{split}
	l_{k,m,i}\left(\boldsymbol{\theta}  \right) =
	T_{k, m, i} \,  
	\ln \lambda_{k, m}(\hat{t}_{k,i}) - 
	R_{k, m}(t_{k,i}) \int_{t_{k,i}}^{ \min\{\hat{t}_{k,i},t_{k,i+1}\} } \lambda_{k, m}(u) \dif u \\
	- 	R_{k, m}(\hat{t}_{k,i}) \int_{ \min\{ \hat{t}_{k,i}, t_{k,i+1}\} }^{t_{k,i+1}}   \lambda_{k, m}(u) \dif u.
\end{split}
\end{equation*} 

\subsection{Pricing formula}
\label{sec:pricing}

We use the estimated transition intensities to derive actuarially fair premiums. Since premiums are paid at discrete points in time in practice, we work in a discrete-time setting. For simplicity, we consider annual time intervals, noting that the framework can be readily adapted to more frequent payment intervals.

A key step in the premium calculation is the computation of multi-year transition probabilities. These are obtained from the one-year transition probabilities via the Chapman-Kolmogorov equations. The one-year transition probabilities, in turn, are computed from the transition intensities using the matrix exponential, which solves the Kolmogorov forward equations.

Actuarially fair premiums are derived by equating the expected present value (EPV) of premiums to the EPV of benefits. The EPV of premiums equals the premium itself in the case of a single lump-sum payment. Alternatively, if a level premium of $P$ is paid annually in advance, and lapses are ignored, the EPV of premiums is given by
\[
\mathrm{EPV}(\text{Premiums}) = P \sum_{t=0}^\infty v^t \Pr(J_{x+t} \in \mathcal{P} \mid J_x = \text{Initial state}),
\]
where $v$ is the discount factor, $J_x$ denotes the health state at age $x$, and $\mathcal{P}$ refers to the set of states in which premiums are payable.

Turning to benefits, a common type of benefit is payable while the insured is in a given state. Suppose that an amount $B_t$ is payable at time $t$, conditional on the insured being in a benefit-paying state. The EPV of benefits is given by
\[
\mathrm{EPV}(\text{Benefits}) = \sum_{t=0}^\infty  v^t \, B_t \, \Pr(J_{x+t} \in \mathcal{B} \mid J_x = \text{Initial health state}),
\]
where $\mathcal{B}$ is the set of states in which benefits are payable. 

Another common type of benefit is a death benefit. If a unit benefit is payable at the end of the year of death, the EPV of the death benefit is
\begin{align*}
& \quad \mathrm{EPV}(\text{Death benefit}) \\ 
&= \sum_{t=0}^\infty  v^{t+1} \, \left[ \Pr(J_{x+t+1} = \text{Dead} \mid J_x = \text{Initial state}) - \Pr(J_{x+t} = \text{Dead} \mid J_x = \text{Initial state}) \right].
\end{align*}

\section{Fair Pricing Methodology}
\label{sec:fairpricing}

Pricing methodologies that incorporate fairness constraints are rarely built from scratch. Instead, they typically begin with a model known to perform well on the desired task and then modify aspects of the model-fitting process to improve specific fairness metrics. These modifications can take place at three points relative to when the model is fit: before (pre-processing), during (in-processing), or after (post-processing). 

In this section, we first outline the most prominent notions of fairness used in practice. We then discuss fair pricing methodologies at each of these three stages of modification: pre-processing, in-processing, and post-processing. Finally, we discuss how the characteristics of the methodologies at each of the three stages constrain generalization to long-term contexts.

\subsection{Notions of fairness}

To discuss existing fair pricing methodologies, we first define commonly used notions of fairness. We use the following notation throughout this discussion:
\begin{itemize}
\item $Y$: the response variable, usually representing the realized loss and used in model estimation. 
\item $\hat{Y}$: the model prediction, usually representing the estimated premium which corresponds to the expected loss. 
\item $\mathbf{Z}$: the vector of covariates representing policyholder characteristics used for underwriting.
\item $S$: the sensitive attribute for the purposes of assessing fairness.  
\end{itemize}

Notions of \emph{group fairness}, developed extensively in the machine learning literature, are central to many existing fair pricing methodologies. The three most commonly used notions are the following \citep{barocas2023fairml}; note that the notation $\perp$ denotes statistical independence:
\begin{itemize}
\item \textbf{Demographic parity:} Also known as \emph{independence}, it requires that $\hat{Y} \perp S$. 
\item \textbf{Equalized odds:} Also known as \emph{separation}, it requires that $\hat{Y}\perp S \mid Y$. 
\item \textbf{Predictive parity:} Also known as \emph{sufficiency}, it requires that $Y\perp S \mid \hat{Y}$. 
\end{itemize}
Both equalized odds and predictive parity require conditioning on $Y$ and $\hat{Y}$, respectively. This conditioning is natural in classification settings, where outcomes ($Y$) and predictions ($\hat{Y}$) are categorical. In actuarial applications, however, both $Y$ and $\hat{Y}$ tend to be continuous variables, making these notions of fairness difficult to utilize. As a result, actuarial fair pricing methodologies that rely on group fairness most often focus on demographic parity \citep{grari2022fairml, hu2024seqfair, lindholm2024proxy}. This paper adopts the same focus.

While useful, enforcing notions of group fairness may result in unintended cross-subsidization in the actuarial context. For this reason, fair pricing methodologies may instead appeal to notions of \emph{individual fairness}. Broadly speaking, individual fairness aims to ensure that similar individuals are treated similarly \citep{dwork2012indfair}. The primary challenge in applying this notion lies in defining an appropriate metric of similarity. Consequently, much of the existing literature on individual fairness proposes intuitive fairness measures that can be viewed as falling within this category, rather than proposing an appropriate metric.

One notion of fairness this paper focuses on is the avoidance of \emph{proxy discrimination} as defined by \citet{lindholm2022discriminationfree}. They notice that, even when the sensitive attribute $S$ is excluded from a model, conditioning on certain values of $\mathbf{Z}$ may yield $\E[Y \mid \mathbf{Z}=\z] \approx \E[Y \mid \mathbf{Z}=\z, S=s]$ for some $s$. In such cases, information about $S$ is inadvertently incorporated through the covariates $\mathbf{Z}$. This phenomenon follows from the tower property of expectation:
\begin{align}\label{eq:lindholmtower}
\E[Y \mid \Z=\z]&= \int_s \E[Y \mid \Z=\z, S=s]\dif \mathbb{P}(S=s \mid \Z=\z).
\end{align}
Proxy discrimination therefore arises from the confluence of two effects: when $S$ has a non-negligible effect on the mean, and when $\mathbf{Z}$ is informative for $S$. One way to mitigate this issue is to replace $\mathbb{P}(S=s \mid \Z=\z)$ with a probability measure that does not depend on $\mathbf{Z}$. This idea underlies the methodology proposed by \citet{lindholm2022discriminationfree}, which we discuss in more detail in Section~\ref{sec:processing}.

\subsection{Types of processing for incorporating fairness}
\label{sec:processing}

We now discuss pre-processing, in-processing, and post-processing methods in turn, using one representative methodology for each to illustrate their key ideas.

\emph{Pre-processing} methods aim to transform input data so that models trained on transformed data can produce predictions that better conform to a desired notion of fairness. A representative example is the optimal transport methodology of \citet{lindholm2024proxy}. Recall that demographic parity requires $S\perp\hat{Y}$. If $\hat{Y} = g(\Z)$ for some $g$, then this requirement is satisfied whenever $\Z\perp S$. This methodology achieves such independence through the following steps:
\begin{itemize}
    \item Identify a candidate probability distribution under which $\Z\perp S$ is satisfied. 
    \item Identify an optimal transport map $r$ to transform the observed distribution of $\mathbf{Z}$ to this candidate probability distribution. 
    \item Defining $\mathbf{Z}^{\perp} := r(\mathbf{Z})$, we have $\mathbf{Z}^{\perp} \perp S$ by construction. Hence, any model trained on $(\mathbf{Z}^{\perp}, Y)$ yields predictions $\hat{Y} = g(\mathbf{Z}^{\perp})$ that satisfy demographic parity with respect to $S$.
\end{itemize}

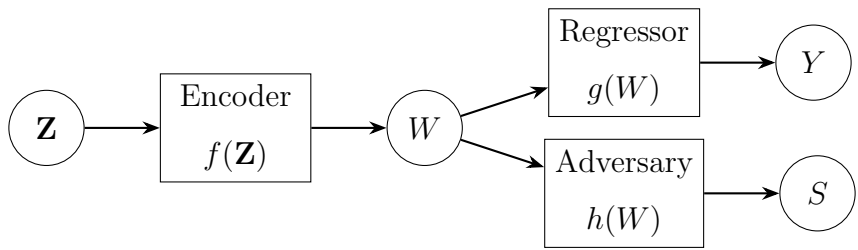
\begin{figure}[ht!]
\centering
\begin{tikzpicture}[
    node distance=1.5cm and 1cm,
    box/.style={draw, rectangle, minimum width=2cm, minimum height=1cm},
    circ/.style={draw, circle, minimum size=1cm},  
    arrow/.style={-Stealth, thick}
]

\node[circ] (Z) {$\Z$};
\node[box, align = center, right= of Z](encoder){Encoder \\ $f(\mathbf{Z})$};
\node[circ, right=of encoder] (W) {$W$};
\node[right = 2cm of W](mid){};
\node[box, align = center, above=0cm of mid] (regressor) {Regressor \\$g(W)$};
\node[circ, right=of regressor] (Y) {$Y$};
\node[box, align = center, below=0cm of mid] (adversary) {Adversary\\$h(W)$};
\node[circ, right=of adversary] (S) {$S$};
 \draw[arrow] (Z) -- (encoder);
 \draw[arrow] (encoder) -- (W);
 \draw[arrow] (W) -- (regressor);
 \draw[arrow] (regressor) -- (Y);
 \draw[arrow] (W) -- (adversary);
 \draw[arrow] (adversary) -- (S);
\end{tikzpicture}
\caption{The adversarial learning framework of \citet{beutel2017data}. \textit{Note}: $W = f(\mathbf{Z})$ refers to the trained representation, whereas $f$, $g$ and $h$ are functions learned by the network. }
\label{fig:madras}
\end{figure}

\emph{In-processing} methods modify an existing model by altering the training objective, the training procedure, or both, so that the resulting predictions conform to a desired notion of fairness. A representative example is the adversarial debiasing methodology of \citet{beutel2017data}, which incorporates both types of modifications. Specifically, this methodology trains a neural network with the structure as seen in Figure \ref{fig:madras}. $W$ is an intermediate representation that is encouraged to contain no information about the sensitive attribute $S$, by training $W$ adversarially to satisfy $W \perp S$. The standard loss of the predictions with respect to $Y$ (e.g., a likelihood-based loss) is then augmented with a penalty corresponding to the \emph{negative} of the accuracy of predicting $S$ from $W$. If the training succeeds in enforcing $W \perp S$, then $\hat{Y} = g(W)$ is also independent of $S$, thereby guaranteeing demographic parity. 

\emph{Post-processing} methods operate by applying an additional adjustment after predictions are generated, producing revised predictions $\hat{Y}$ that adhere to a desired notion of fairness. A representative example is the methodology of \citet{lindholm2022discriminationfree}. Recall from Equation~\eqref{eq:lindholmtower} their observation that $\E[Y \mid \Z=\z]$ implicitly incorporates information on $S$ through the tower property of expectation. To remove this implicit dependence and thereby avoid proxy discrimination, they propose replacing $\mathbb{P}(S=s|\mathbf{Z}=\mathbf{z})$ with a probability measure that does not depend on $\mathbf{Z}$. Their approach proceeds in two steps:
\begin{itemize}
\item Fit a model that predicts $\mu(\z, s) = \E[Y \mid \Z=\z, S=s]$. 
\item Obtain the discrimination-free mean functional as
\begin{align*}
\mu^*(\z)&= \int_s \mu(\z, s) \dif \mathbb{P}^*(S=s), 
\end{align*}
where $\mathbb{P}^*$ is a user-specified probability measure.
\end{itemize}

\subsection{Challenges in adapting existing methodologies to long-term insurance}

We have seen that in-processing and post-processing techniques for incorporating fairness constraints rely on a single, well-defined outcome variable. However, in the long-term insurance setting where transition rates are estimated via multi-state models, it is unclear which quantity, if any, could serve this role. Over the lifetime of an insured, transitions may occur between multiple states, and in some settings, such as long-term care, individuals may return to a previously attained state (e.g., through recovery from disability). As a result, the notion of a single representative outcome is not well defined.

More broadly, the difficulty in choosing a single representative outcome complicates both the assessment of fairness and the formulation of fair pricing methodologies. For instance, demographic parity requires independence between the model predictions and the sensitive attribute. When it is unclear what constitutes the relevant prediction in a long-term context, enforcing demographic parity becomes problematic, even for methods that ostensibly do not interact with the response variable. Take the pre-processing methodology of \citet{lindholm2024proxy} as an example. Although it operates solely by transforming covariates to be independent of the sensitive attribute, such transformations are difficult to evaluate when it is unclear whether the resulting predictions actually satisfy demographic parity.

One may argue for the expected present value of the product’s cash flows to serve as the outcome. However, this choice renders the outcome product-dependent and does not align with standard practice in multi-state modeling, where the primary quantities being estimated are transition rates rather than prices. This contrasts with short-term insurance products, where the outcome variable directly corresponds to the expected loss being modeled. In general, whether the outcome used should be product-dependent (e.g., price) or product-agnostic (e.g., transition rates) needs to be resolved in developing a coherent fair pricing methodology for long-term insurance products. 

\section{A Unified Framework for Multi-state Modeling}
\label{sec:unified}

The primary difficulty in extending fair pricing methodology to long-term insurance products lies in the absence of a clearly defined outcome variable within multi-state models. In this section, we propose a unified framework that reformulates any multi-state modeling problem as a set of Poisson regressions. In this setting, both the outcome variable and the corresponding predictions are well defined, thereby enabling the adaptation of fair pricing methodologies to the long-term insurance setting. 

We begin by outlining the Poisson-survival connection, which forms the backbone of this framework. We then discuss practical considerations for implementing this approach, with particular emphasis on actuarial applications.

\subsection{The Poisson-survival connection}

We introduce our unified framework by building upon the notations used in Section~\ref{sec:multistate}, but in a simplified setting involving a single individual. The key to this unified framework is the fact that any survival process can be viewed as a (possibly non-homogeneous) stopping Poisson process. Thus, over any fixed time interval, the number of events generated by such a process follows a Poisson distribution.

This representation extends to multi-state models. The time spent in a given state may be viewed as a competing risks process, where the competing risks correspond to the $M$ possible transitions. Each of the $M$ transitions can be modeled as an independent stopping Poisson process. As a result, over a fixed time interval, the number of occurrences for each of the $M$ transitions follows $M$ independent Poisson distributions. 

This connection becomes explicit when examining the likelihood for a single individual. For simplicity, suppose each of the $M$ types of possible transitions has a constant transition rate $\lambda_m$, and that all transitions are possible from the individual's current state. Let $\hat{t}$ denote the time of transition, and let $T_m$ be an indicator taking value 1 if the observed transition is of type $m$ and 0 otherwise. Let $\bm{\theta}$ denote the vector of model parameters. The likelihood is given by
\begin{align*}
    L(\bm{\theta})&= \exp \left( -\hat{t}\sum_{m=1}^M \lambda_m \right)\prod_{m=1}^M \lambda_m^{T_m}.
\end{align*}
We may rewrite this likelihood, up to a proportionality constant, as
\begin{align*}
    L(\bm{\theta})\propto \prod_{m=1}^M\frac{ \exp \left( -\hat{t} \cdot  \lambda_m \right)(\hat{t} \cdot  \lambda_m)^{T_m}}{T_m!}, 
\end{align*}
which is equivalent to the likelihood of $M$ independent Poisson processes with intensity $\lambda_m$ observed over an interval of length $\hat{t}$, where the observed count for each process, $T_m$, is limited to either 0 or 1.

In a practical multi-state model, an individual does not remain in a single state indefinitely. Because Poisson processes have independent increments over non-overlapping time intervals, the time spent in each successive state may be modeled as independent of the time spent in other states. Moreover, although there are $M$ possible transitions in the overall multi-state model, only a subset is possible from any given state. Thus, for each sojourn in a particular state, the competing risks component reduces to those transitions that are possible from that state. 

\subsection{The unified framework in practice}
\label{sec:unifiedpractice}

The assumption that each $\lambda_m$ is constant, both over an individual’s lifetime and across individuals, is overly restrictive for practical applications. We therefore relax this assumption by allowing $\lambda_m$ to vary with covariates. Time-varying covariates are readily accommodated by the model, as they are commonly specified in piecewise constant form in practice. The model can thus be partitioned into intervals over which the transition rates remain constant.

In long-term insurance, transition rates are traditionally modeled using a log-linear form, as given by Equation~\eqref{eq:intensity}. However, several fair pricing methods explicitly require the use of neural networks to model the predicted value (see Section~\ref{sec:inprocessing}), motivating a more flexible methodology. By viewing multi-state models as a set of Poisson regressions, one may employ any method compatible with a Poisson likelihood loss, such as generalized additive models, tree-based ensemble methods, and, importantly, neural network-based methods commonly used in fair pricing literature. This leads to the more general formulation
\begin{equation} \label{eq:intensity2}
\ln \lambda_{k, m}(t) =
	f_m(\mathbf{z}_k, \boldsymbol{\omega}_k(t)).
\end{equation}

We now specialize the framework to align with standard actuarial pricing practice. Insurers typically employ tables of annual transition probabilities for pricing, indexed by the insured's attained age. We accommodate this convention by assuming a constant $\lambda_m$ for each integer-valued age $x$, where age may be defined as age last birthday, age nearest birthday, or age next birthday. For the rest of this paper, we measure age as age last birthday. 

We next consider the covariate structure relevant for underwriting and pricing. Because transition rates are estimated for underwriting purposes and applied at policy inception, covariates other than age are treated as static for a given insured. Accordingly, the $k$\textsuperscript{th} insured is associated with a vector of static covariates $\mathbf{z}_k$. We further assume that the transition rates depend on time only through the insured's attained age. Under these assumptions, the transition rate for insured $k$,  transition $m$ at age $x$, denoted $\lambda_{k, m}(x)$, is modeled via the following specialization of Equation~\eqref{eq:intensity2}:
\begin{align*}
\ln \lambda_{k, m}(x)&= f_m(\mathbf{z}_k, x).
\end{align*}
With this structure in place, for $(k, m, x)$ indexing the entry for age $x$ for insured $k$ in transition $m$, we model the the transition event $T_{k, m, x}$ as a Poisson random variable with mean $\tau_{k, m, x}\cdot \lambda_{k, m}(x)$, where $\tau_{k, m, x}\in [0, 1]$ denotes the exposure associated with $T_{k, m, x}$. 

The value of $\tau_{k, m, x}$ depends on how the health state of insured $k$ changes at age $x$.  Recall that only a subset of transitions may be possible from any given health state. Thus, if throughout age $x$, insured $k$ remains in states from which transition $m$ is impossible, we {set $\tau_{k, m, x}=0$ and exclude the observation from the estimation sample}; otherwise, $\tau_{k, m, x}>0$. 

Running the Poisson regression requires the data to be structured so that age remains constant on each interval and that each transition count is paired with the appropriate exposure. An illustration of this restructuring process, using the data introduced in the case study of Section~\ref{sec:casestudy} and its associated multi-state model, is provided in Appendix~\ref{sec:transformations}. Once this restructuring is done, the Poisson regression for transition $m$ can then be run using the log-link function, with $T_{k, m, x}$ as the number of occurrences, i.e., the response variable, $\ln \tau_{k, m, x}$ as an offset variable, and $(\mathbf{z}_k, x)$ as covariates.

\section{Case Study: Fair Pricing for LTCI}\label{sec:casestudy}

With the unified framework of Section~\ref{sec:unified} in place, we now introduce the general framework for applying fair pricing methods to long-term insurance products. Since this involves restructuring the data to accommodate the Poisson regressions required under the unified framework, the framework is best illustrated through an example. We conduct a case study that prices a stylized LTCI product using the HRS dataset. We focus on LTCI because both its underlying multi-state model and benefit structure are sufficiently complex to highlight the practical challenges that may arise when implementing this methodology. 

In this section, we first describe how we use the HRS data in our case study. We then present three types of LTCI premiums, including a fairness-adjusted premium obtained by adapting the discrimination-free pricing formula of \citet{lindholm2022discriminationfree}.

\subsection{Dataset construction from the HRS}
\label{sec:hrsdata}

Given the proprietary nature of LTCI portfolio data, we rely on publicly available data from the HRS for our case study. The HRS is a biennial panel survey of initially non-institutionalized Americans aged 50 and older since 1992. The extensive range of variables collected by HRS and its longitudinal design make this dataset well-suited for modeling health state transitions and examining their associations with various covariates. We analyze data from 1998 to 2020 due to inconsistencies in survey questions on functional limitations before 1998 \citep{fong2015disaggregating}. To reduce the likelihood of multiple unobserved health state transitions between interview waves, we restrict our sample to respondents who participated in every interview following their entry into the survey.

We construct two datasets to support our modeling approach. The first dataset records the timing of health state transitions, assuming the underlying process follows the three-state model shown in the right panel of Figure~\ref{fig:multistateModel}. The health states are defined based on six Activities of Daily Living (ADLs) and the Langa-Weir classification of cognitive function \citep{langa2020langaweir}. Consistent with benefit triggers commonly used in commercial LTCI products \citep{hhs2020receiving}, an individual is classified as disabled if they have two or more ADL limitations or exhibit cognitive impairment. The date of death is directly available in the HRS, and all other transitions are assumed to occur at the midpoint between two consecutive interviews. 
For estimation, the raw transition data from the first dataset are not directly suitable for the Poisson regressions required by the unified framework. We use the procedure described in Appendix~\ref{sec:transformations} to reshape the data into the form discussed in Section~\ref{sec:unifiedpractice}.

The second dataset focuses on covariate information collected at each individual's initial interview, mirroring the underwriting process by recording baseline characteristics only at study entry. Table~\ref{tab:list_of_vars} lists the covariates included in our study, all of which are based on self-reported data. These covariates include factors likely to influence disability and mortality, such as health behaviors (e.g., drinking and smoking), socioeconomic variables (e.g., education), as well as demographic attributes relevant to fairness considerations, such as race (with three categories: White/Caucasian, Black/African American, and Other) and gender. 

\begin{sidewaystable}[htbp]
    \centering
    \caption{List of variables. C denotes categorical variables, and N denotes numerical variables.}
    \label{tab:list_of_vars}
    \begin{threeparttable}	
    \begin{tabular}{l l l}
    \toprule
    Demographic & Health and disease & Finance \\
    \midrule
    Age (N) & Self-rated health (C) & Non-housing wealth\tnote{\S} (N)\\ 
    Gender (C) & Ever had diabetes (C) & Household income\tnote{$\|$} (N)\\ 
    Race\tnote{\textdagger} (C)  & Ever had lung disease (C) & \\ 
    Ethnicity\tnote{\textdaggerdbl} (C) & Ever had heart problems (C) & \\ 
    Years of education (C) & Ever had stroke (C) & \\ 
    Census region (C) & Body mass index (BMI) (N) & \\ 
    Marital status (C) & Ever drinks any alcohol (C)  & \\ 
    Labor force participation (C)  & Ever smoked cigarettes (C)  & \\ 
    Number of household members (N)& & \\ 
    \bottomrule
    \end{tabular}
    \begin{tablenotes}
        \item[\textdagger] Race includes three categories: White/Caucasian, Black/African American, and Other.
        \item[\textdaggerdbl] Ethnicity indicates whether the survey respondent identifies as Hispanic.
        \item[\S] Non-housing wealth includes both tangible assets (e.g., vehicles, businesses) and financial assets (e.g., retirement accounts, stocks, bonds, savings), net of non-mortgage debt. 
        \item[$\|$] Household income is the total income for the previous calendar year, including respondent and spouse earnings, pensions and annuities, Supplemental Security Income and Social Security Disability, Social Security retirement benefits, unemployment and workers' compensation, other government transfers, household capital income, and other income sources.
        \item[] \textit{Note}: Job physicality was initially included but was later excluded due to a high proportion of missing values.
    \end{tablenotes}		
\end{threeparttable}
\end{sidewaystable}

Before the covariate dataset can be used, missing values must be addressed. We first remove variables with a particularly large proportion of missing data, which leads to the exclusion of job physicality, as it contains 14\% missing values. We then delete all remaining incomplete observations, resulting in the removal of 3\% of entries and yielding a complete final dataset.

\subsection{LTCI pricing under different settings}
\label{sec:LTCIpricing}

Although the HRS is not designed as an insurance dataset, the health state transition rates estimated from the surveyed population are sufficient for our purposes and can be used to price LTCI products. We model transition rates using log-linear models, as these are widely adopted in the literature. This approach allows us to leverage the {\tt glm} routine in the R programming language with a log-link and the Poisson family of distributions. When the data is appropriately structured, estimation proceeds straightforwardly using R’s native formula and offset interface.

To develop a parsimonious model, we fit the model with all the covariates and then identify covariate transformations---such as regrouping categorical levels---that do not materially affect model fit. As this case study is intended to showcase our pricing methodology, we relegate these details to Appendix~\ref{app:simplifications}. After these transformations are implemented, the model is refit using all available covariates.

We treat the transition rates estimated from our model, for the purposes of this case study, as if they describe the experience of a representative LTCI portfolio. To approximate a realistic insured population, we restrict our pricing analysis to individuals aged 50--80 at issue (on an age-last-birthday basis) and exclude those with disqualifying health conditions. Specifically, eligible individuals must not be disabled at issue and must not report severe heart or lung conditions, diabetes, or stroke. These restrictions yield a relatively healthy baseline population suitable for LTCI pricing analysis.

We consider an LTCI product with a lump-sum premium paid at policy inception and an annual benefit of \$1 for each year spent in the disabled state. We assume that each individual in our baseline insured population represents an insured whose coverage begins immediately after the individual's previous birthday. We also assume that this insured's transition rates are given by those predicted for the same individual, with a terminal age of 110 imposed on the transition rates. Using the pricing formula introduced in Section~\ref{sec:pricing}, the lump-sum premium for a policy issued to an individual aged $x$ is given by
\[
\text{Lump sum premium} = \sum_{t = 0}^{110 - x} v^t \Pr(J_{x+t} = \text{Disabled} \mid J_x = \text{Healthy}),
\]
where the discount factor $v$ is set to $1.03^{-1}$. 

We choose race as the sensitive attribute, denoted by $S$. All other covariates, where we denote the non-age covariates collectively as $\mathbf{z}$, are listed in Table~\ref{tab:list_of_vars}. To evaluate fairness in pricing, we compute three types of premiums under different modeling assumptions:
\begin{itemize}
    \item \textbf{Best-estimate price}: Transition rates $\hat{\lambda}_m(\mathbf{z}, x, s)$, for transition type $m = 1, \ldots, 4$, are estimated using all available covariates $\mathbf{z}$, the policyholder's age $x$, and their sensitive attribute $s$, based on the fitted multi-state GLM model:
    \[
    \hat{\lambda}_m(\mathbf{z}, x, s) = \exp\!\left( \hat{f}^{(\mathrm{Best})}_m(\mathbf{z}, x, s) \right),
    \]
    where $\hat{f}^{(\mathrm{Best})}_m(\cdot)$ denotes the fitted GLM predictor function that includes race as an explanatory variable.
    \item \textbf{Race-blind price}: Transition rates $\hat{\lambda}_m(\mathbf{z}, x)$ are re-estimated after removing race from the covariate set, which assumes the insurer is unaware of the sensitive attribute:
    \[
    \hat{\lambda}_m(\mathbf{z}, x) = \exp\!\left( \hat{f}^{(\mathrm{Blind})}_m(\mathbf{z}, x) \right),
    \]
    where $\hat{f}^{(\mathrm{Blind})}_m(\cdot)$ denotes the re-fitted GLM predictor function excluding race from both data and estimation.
    \item \textbf{Fairness-adjusted price}: Transition rates are adjusted using the $\lambda^*_m(\mathbf{z}, x)$ obtained using the procedure which will be described in detail in Section~\ref{sec:postprocessing}.
\end{itemize}

\subsection{Tailored post-processing for long-term insurance}
\label{sec:postprocessing}

To obtain fairness-adjusted transition rates, we adapt the discrimination-free pricing methodology of \citet{lindholm2022discriminationfree} to the long-term insurance context.
The objective of this methodology is to prevent nonsensitive covariates from acting as proxies for the sensitive attribute. In our setting, these covariates are $(\mathbf{Z}, x)$. Proxying through them can therefore be mitigated by fitting each $\lambda_m$ as a function of $(\mathbf{Z}, x, S)$, and then taking an appropriate expectation of $\lambda_m$ with respect to a chosen distribution for $S$. 

A standard choice for this distribution is the empirical marginal distribution. However, this raises the question of which empirical distribution to use: that for the sample of policies, or the respective ones for each Poisson regression problem. As discussed briefly in Sections~\ref{sec:unifiedpractice} and~\ref{sec:hrsdata} and in detail in Appendix~\ref{sec:transformations}, additional observations are artificially created to maintain constant transition rates over each time interval. The number of such observations is an artifact of each insured’s realized health trajectory. Hence, the empirical distribution at the policy level is the more appropriate choice. We therefore propose the following post-processing procedure for long-term insurance:
\begin{enumerate}
    \item Begin with the sample of observed policies. Let $\mathbf{z}_k$, $x_{u, k}$, and $s_k$ denote the vector of covariates, the age at entry, and the sensitive, respectively, for the $k$\textsuperscript{th} insured, and let $\mathbb{P}_n$ denote the empirical distribution of this sample.
    
    \item Construct $M$ datasets corresponding to each type of transition. Using the Poisson regression technique of choice, estimate $\lambda_m$ as a function of $\mathbf{Z}, x$, and $S$.
    
    \item For each desired combination of $(\mathbf{z}, x)$, calculate the discrimination-free transition rate as
    \begin{align*}
    \lambda_m^* (\mathbf{z}, x)&= \int \hat{\lambda}_m (\mathbf{z}, x, s)\dif \mathbb{P}_n(s)
    \end{align*}
    
    \item The price for any insured $P(\mathbf{z}, x_u)$ can then be calculated using  $\hat{\lambda}^*_m(\mathbf{z}, x)$.  
\end{enumerate}

This procedure ensures that the calculated $\lambda^*(\mathbf{z}, x)$ is estimated free from proxy discrimination with respect to $(\mathbf{Z}, x)$, and consequently that the resulting price $P(\mathbf{z}, x_u)$ inherits this property.

\subsection{Results and discussion }

\begin{figure}[ht!]
\centering
\includegraphics[width = \textwidth]{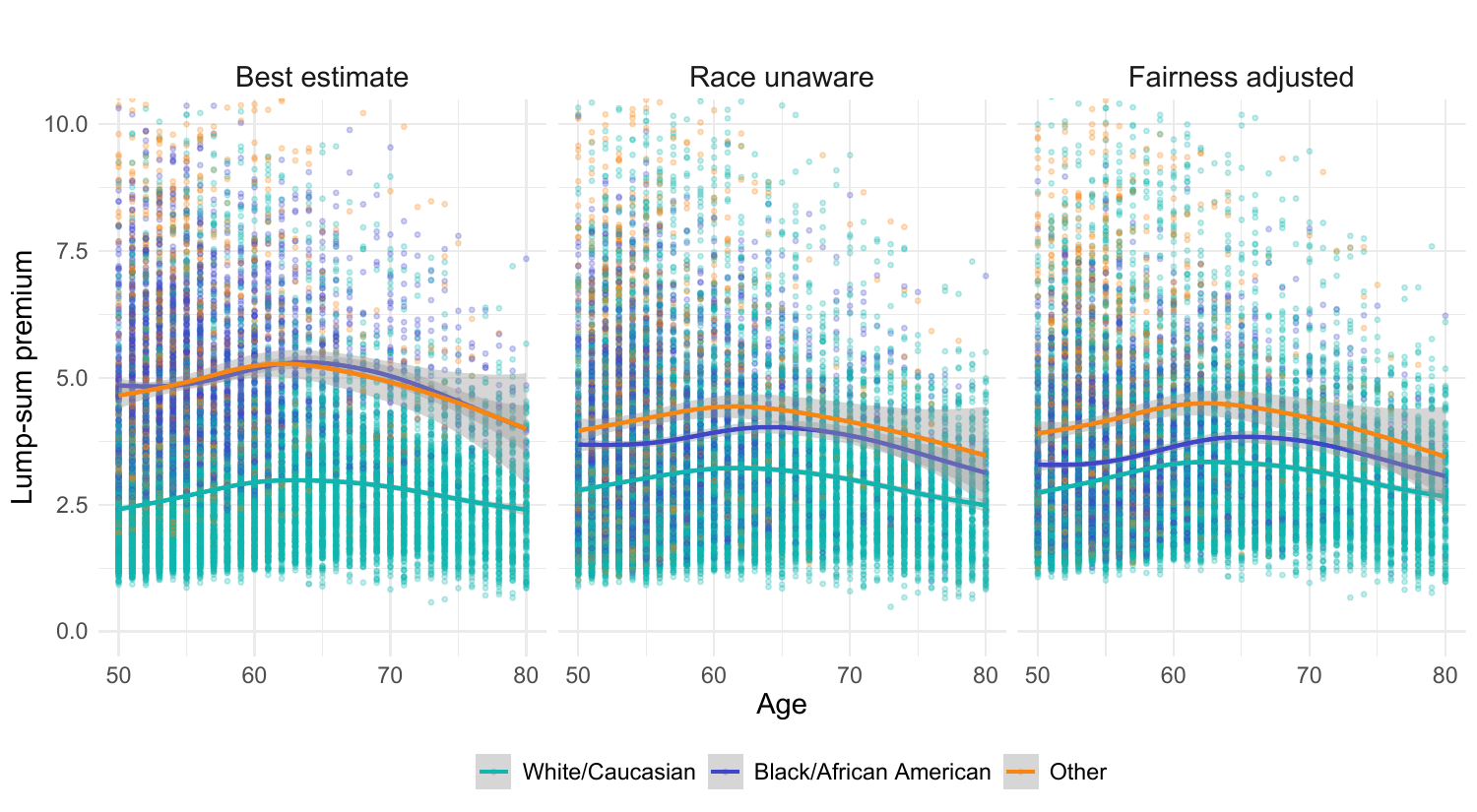}
\caption{Lump-sum premiums by racial group. \textit{Note}: Each panel represents a different modeling assumption, as indicated by the panel title. There are three smoothed lines in each panel, one for each racial group. Each smoothed line is generated using a generalized additive model (GAM), regressing the lump-sum premium on age. The gray band represents the 95\% confidence interval of the GAM.}
\label{fig:lump_sum_glm_unaware_lam_exp}
\end{figure}

Figure~\ref{fig:lump_sum_glm_unaware_lam_exp} shows premiums computed under three different modeling assumptions. Each point represents the lump-sum premium charged to an individual. For a given age and racial group, premiums vary due to differences in nonsensitive covariates used for pricing. To facilitate comparison across racial groups, we overlay three smoothed curves in each panel, one for each racial group. These curves are obtained using generalized additive models, with the lump-sum premium regressed on age.

Interestingly, across races and modeling assumptions, the smoothed lump-sum premiums exhibit relatively little variation with age. This pattern reflects the interaction of two opposing effects. As age increases, the probability of transitioning from healthy to disabled rises, while that of recovery from disability declines, increasing the expected present value of benefits. At the same time, mortality rates increase with age, reducing the expected duration an individual spends in the disabled state.

We now visually assess how each set of prices compares in terms of group fairness. We first compare the best estimate prices, whose underlying transition rates explicitly include race as a covariate, with the race-unaware prices, which omit the covariate of race. The clearer separation between the White/Caucasian and the other two groups under the best estimate prices suggests that race is an important predictor in the fitted model. Indeed, according to the likelihood contributions for each predictor in Appendix~\ref{app:likelihood}, race is the third most important predictor in the model, behind age and years of education. 

Although race-unaware prices exclude race as a covariate, the gap between the average prices nevertheless persists, albeit to a smaller extent. While race-unaware transition rates no longer include direct effects of race, indirect effects may still arise through other covariates that are correlated with race. Such proxy effects likely contribute to the remaining disparity.

To address this issue, we apply the post-processing procedure adapted from \citet{lindholm2022discriminationfree} to ensure that such inference of the sensitive attribute is no longer present in the adjusted transition rates. However, as shown by \citet{lindholm2024proxy}, the absence of proxy discrimination and notions of group fairness in general do not imply one another. Hence, we do not expect price differences to be fully eliminated. Indeed, a comparison of the middle and right panels of Figure~\ref{fig:lump_sum_glm_unaware_lam_exp} shows that, although the gap persists, it is further reduced. Since the reweighting method eliminates indirect effects of race, we interpret the remaining gap as reflecting average differences attributable solely to the direct effects of the other covariates. 

Notably, the reduction in the price gap is most pronounced between the Black/African American and White/Caucasian groups, whereas the gap between the Other and White/Caucasian groups appears largely unchanged. This suggests that for the Black/African American group, the race-unaware prices contain substantial indirect effects of race through other covariates, which are effectively mitigated by the adjusted prices. In contrast, indirect effects of race appear to be much weaker for the Other group.

\section{A Fair Pricing Framework for Long-Term Insurance}\label{sec:longtermfair}

Through the case study, we illustrated how to apply our fair pricing framework in the LTCI context using the post-processing methodology of \citet{lindholm2022discriminationfree}. In this section, we provide a pipeline for applying this framework more broadly. We first describe the overall framework and then provide two illustrations---corresponding to pre-processing and in-processing approaches---to demonstrate its practical implementation and to discuss issues that may arise in application.

\subsection{General framework}

The unified framework introduced in Section~\ref{sec:unified} expresses transition rate estimation in multi-state models as a set of Poisson regression problems. This leads to a two-step procedure for applying fair pricing methods to long-term insurance products: 
\begin{enumerate}
    \item Reformulate the multi-state modeling problem in terms of its constituent Poisson regression problems.
    \item Apply the desired fair pricing method to each of the constituent Poisson regression problems.
\end{enumerate}
This two-step procedure has proven effective for the post-processing approach of \citet{lindholm2022discriminationfree}, as demonstrated in Section~\ref{sec:casestudy}. However, as we shall see in Sections~\ref{sec:preprocessing} and~\ref{sec:inprocessing}, extending fairness concepts to the multi-state setting requires particular care because age enters not only into the estimation of transition rates, but also into the calculation of premiums through the age at underwriting. Consequently, care needs to be taken in the treatment of age in each case, depending both on the exact methodology being considered and the notion of fairness desired.

\subsection{Illustration 1: Pre-processing via optimal transport}
\label{sec:preprocessing}

We begin by illustrating how our framework applies to the optimal transport methodology of \citet{lindholm2024proxy}, which performs pre-processing. Optimal transport maps are generally ill-defined for categorical variables, which lack a natural distance metric. This is an issue even for short-term insurance products. For simplicity, we present the case only where $\mathbf{Z}$ is solely composed of continuous covariates. 

Recall that demographic parity requires independence between the outcome variable and the sensitive variable, i.e., $\hat{Y}\perp S$. In \citet{lindholm2024proxy}, this requirement is enforced by altering covariate values such that they are independent of the sensitive variable $S$. In the long-term insurance setting, the main challenge concerns the treatment of age, and the appropriate handling depends on the form of demographic parity being imposed.

Let $x_u$ denote the age of an insured at underwriting, and let $P(\mathbf{z}, x_u)$ denote the price charged to an insured with covariates $\mathbf{z}$ and age $x_u$. One could argue that demographic parity should require the prices charged to each insured, $P(\mathbf{Z}, x_u)$, to be independent of $S$, where $\mathbf{Z}$ and $x_u$ are both random realizations from the data. However, enforcing this condition would require altering ages, raising additional fairness concerns. We therefore adopt a simpler requirement: for each fixed value of $x^*$, $P(\mathbf{Z}, x^*)$ should be independent of $S$. 

We propose the following modified procedure for the long-term insurance context:
\begin{enumerate}
    \item Begin with the sample of observed policies, where $\mathbf{z}_k$ and $x_{u, k}$, respectively, denote the vector of covariates and the age at entry of the $k$\textsuperscript{th} insured, and let $\mathbb{P}_n$ denote the empirical distribution of this sample.
    
    \item Identify a candidate distribution $\mathbb{P}^{\perp}$ under which $\mathbf{Z}\perp S$, while preserving the marginal distribution of $x_u$ from $\mathbb{P}_n$.
    
    \item Determine the optimal transport map $r$ which maps the empirical probability distribution $\mathbb{P}_n$ to $\mathbb{P}^{\perp}$.
    
    \item Define $\mathbf{z}_k^{\perp} := r(\mathbf{z}_k)$. Using the modified data $(\mathbf{z}^{\perp}_k, x_{u, k})$, construct the Poisson regression datasets for each of the $M$ transitions, and estimate the corresponding transition rates $\lambda_m$. 
    
    \item Calculate the price for the $k$\textsuperscript{th} insured, $P(\mathbf{z}_k^{\perp}, x_{u, k})$, using the estimated $\lambda_m$. 
\end{enumerate}

While the actual premium charged to each insured still depends on the insured's age at entry, the form of demographic parity guaranteed by this procedure is \emph{conditional on issue age}. For each fixed age $x$ and transition $m$, $\lambda_m(\mathbf{z}_k^{\perp}, x)$ is independent of $S$. The price $P(\mathbf{z}_k^{\perp}, x^*)$ is then a function of $\lambda_m(\mathbf{z}^{\perp}_k, x)$ evaluated over ages $x=x^*, x^*+1, \ldots, 110$ and transitions $m=1, 2, \ldots, M$. Since each of these quantities is independent of $S$, the resulting price is likewise independent of $S$ for each $x^*$.

\subsection{Illustration 2: In-processing via adversarial debiasing}\label{sec:inprocessing}

We now illustrate how the methodology of \citet{beutel2017data} can be adapted to long-term insurance pricing. To this end, we supplement the general description in Section~\ref{sec:processing} with the specific adversarial learning procedure as described below.

The objective function of the model is given by
\begin{align}\label{eq:madrasloss}
\min_{f,g}\;\max_{h}\;
\mathbb{E}\Big[
    L_Y\!\left(
        Y,\;
        g\big(f(\mathbf{Z})\big)
    \right)
    - \alpha\,
    L_S\!\left(
        S,\;
        h\big(f(\mathbf{Z})\big)
    \right)
\Big],
\end{align}
where $L_Y$ is the loss with respect to the outcome variable (here, the likelihood loss), and $L_S$ is the cross-entropy loss. Training proceeds via alternating two steps: an adversarial gradient step to maximize Equation~\eqref{eq:madrasloss} with respect to $h$, and a model gradient step to minimize Equation~\eqref{eq:madrasloss} with respect to $(f, g)$.    

By alternating between the adversary and the model, one can in principle obtain $W \perp S$, while ensuring $W$ remains maximally predictive for $Y$. As in the case of pre-processing,  $W\perp S$ implies $\hat{Y} = g(W) \perp S$, yielding demographic parity. In practice, however, achieving demographic parity often requires sacrificing accuracy to an unacceptable degree and vice versa. The parameter $\alpha$ therefore serves as a tuning parameter to find a suitable balance between these two objectives. 

We now describe how this methodology can be tailored to the long-term context. A key modeling choice concerns the treatment of age, specifically whether it forms part of the representation. Since each $\lambda_m$ is a function of $(\mathbf{Z}, x)$, one possibility would be to include age directly in the representation as $W = f(\mathbf{Z}, x)$. However, it is difficult to define how the adversary should interact with such a representation in a way that guarantees some form of independence between the prices and $S$. We therefore restrict the representation to $W = f(\mathbf{Z})$, and train the network such that $W\perp S$ under the distribution of observed policyholders; the predicted transition rate is then obtained as $\hat{\lambda}_m = g_m(W, x)$. While this specification does not guarantee that $P(\mathbf{Z}, x_u)$ is independent of $S$, it does guarantee that $P(\mathbf{Z}, x^*)$ is independent of $S$ for each fixed value of $x^*$. 

Given that we are training $M$ regression problems---one for each transition---we again have two choices in how the representations are trained. The more straightforward approach is to obtain separate representations and adversaries for each transition. While this approach aligns closely with the divide-and-conquer structure of the unified framework, it is relatively data-inefficient for training the adversary since we simply apply the methodology of \citet{beutel2017data} $M$ times. We therefore relegate this approach to Appendix~\ref{app:madras}.

\begin{figure}[ht!]
\centering
\begin{tikzpicture}[
    node distance=1.5cm and 1cm,
    box/.style={draw, rectangle, minimum width=2cm, minimum height=1cm},
    circ/.style={draw, circle, minimum size=1cm},  
    arrow/.style={-Stealth, thick}
]

\node[circ] (X) {$\Z, x$};
\node[circ, right = of X] (Z) {$\mathbf{Z}$};
\node[box, align = center, right= of Z](encoder){Encoder \\ $f(\mathbf{Z})$};
\node[circ, above = 1cm of Z] (x) {$x$};
 \node[circ, right=of encoder] (W) {$W$};
 \node[circ, right=of W, yshift = -0.1cm] (Wx) {$W, x$};
 \node[box, align = center, below=0.5cm of W] (adversary) {Adversary\\$h(W)$};
 \node[circ, below=0.5cm of adversary] (S) {$S$};
 \node[box, align = center, right=1.0cm of Wx] (regressor2) {Regressor 2 \\$g_2(W, x)$};
 \node[circ, right=of regressor2] (Y2) {$Y_2$};
 \node[box, align = center, above=0.5cm of regressor2] (regressor1) {Regressor 1 \\$g_1(W, x)$};
 \node[circ, right=of regressor1] (Y1) {$Y_1$};
 \node[below=0.5cm of regressor2] (regressordots) {$\vdots$};
 \node[box, align = center, below=0.5cm of regressordots] (regressorM) {Regressor $M$ \\$g_M(W, x)$};
 \node[circ, right=of regressorM] (YM) {$Y_M$};   

  \draw[arrow] (X) -- (Z);
  \draw[arrow] (Z) -- (encoder);
  \draw[arrow] (X) -- (x);
  \draw[arrow] (encoder) -- (W);
  \draw[arrow] (W) -- (Wx);
  \draw [arrow] (x) to [out=0,in=135] (Wx);

  \draw[arrow] (Wx) -- (regressor2.west);
  \draw[arrow] (regressor2) -- (Y2);
  \draw[arrow] (Wx) -- (regressor1.west);
  \draw[arrow] (regressor1) -- (Y1);
  \draw[arrow] (Wx) -- (regressorM.west);
  \draw[arrow] (regressorM) -- (YM);
  \draw[arrow] (W) -- (adversary);
  \draw[arrow] (adversary) -- (S);
\end{tikzpicture}
\caption{Adversarial learning framework tailored to long-term insurance, adapted from \citet{beutel2017data}. \textit{Note}: $\mathbf{Z}$ and $x$ refer to the non-age covariates and age, respectively; $W=f(\mathbf{Z})$ is the trained representation, and functions $f$, $h$ and $g_1, \ldots, g_M$ are functions learned by the network.}
\label{fig:madras2}
\end{figure}
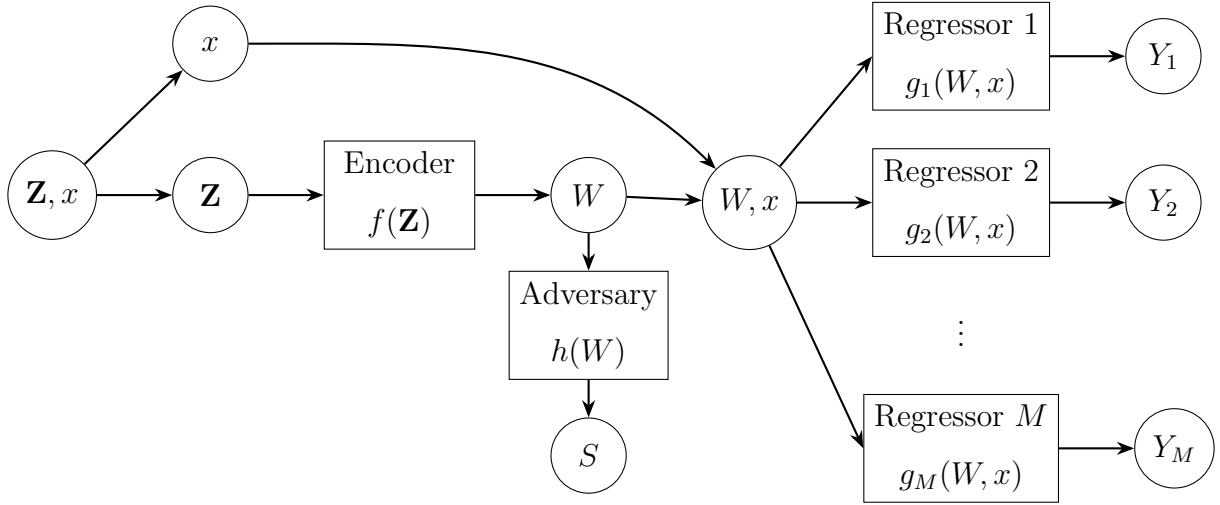

Instead, we focus on a more data-efficient approach which utilizes a common representation and a common adversary across all transitions. Separate regressors for each transition then make use of this common representation, along with age, to obtain predictions. The architecture for this approach is shown in Figure~\ref{fig:madras2}.

The proposed procedure is as follows:
\begin{enumerate}
    \item Begin with the sample of observed policies, with $\mathbf{z}_k$, $x_{u, k}$, and $s_k$, respectively, denoting the vector of covariates, age at entry, and sensitive attribute of the $k$\textsuperscript{th} insured.
    
    \item Let $w_k = f(\mathbf{z}_k)$. Using cross-entropy loss $L_S$, form the adversary loss
    \begin{align*}
    \mathit{Loss}_{Adv}&= \frac{1}{n}\sum_{k=1}^nL_S(s_k, h(w_k)).
    \end{align*}
    
    \item Construct $M$ datasets corresponding to each transition. Let $(k, m, j)$ index the $j$\textsuperscript{th} entry for the $k$\textsuperscript{th} insured and the $m$\textsuperscript{th} transition, with $n_{k, m}$ denoting the number of entries and $n_m = \sum_{k=1}^n n_{k, m}$. Using the Poisson likelihood loss $L_Y$, form the loss for the $m$\textsuperscript{th} transition
    \begin{align*}
    \mathit{Loss}_m &= \frac{1}{n_m}\sum_{k=1}^n \sum_{j=1}^{n_{k, m}}L_Y(Y_{k, m, j}, g_m(w_k, x_{k, m, j})).
    \end{align*}
    
    \item Form the total loss
    \begin{align}\label{eq:madrasloss2}
    \mathit{Loss} &= \frac{1}{\sum_{m=1}^M n_m}\sum_{m=1}^M n_m \mathit{Loss}_m - \alpha \mathit{Loss}_{Adv}.
    \end{align}
    
    \item Alternate between gradient steps to maximize Equation~\eqref{eq:madrasloss2} with respect to $h$, and to minimize Equation~\eqref{eq:madrasloss2} with respect to $(f, g_1, \ldots, g_M)$, until convergence.
    
    \item Calculate the price for the $k$\textsuperscript{th} insured, $P(\mathbf{z}_k, {x_{u, k}})$, using $\hat{\lambda}_m(\mathbf{z}, x) = g_m(f(\mathbf{z}), x)$. 
\end{enumerate}

Under this adversarial learning procedure, $W$ is independent of $S$. For each transition $m$ and fixed age $x$, $\hat{\lambda}_m(\mathbf{z}_k, x) = g_m(W, x)$ is a function of $W$ and is therefore independent of $S$. Thus, as a function of such quantities, $P(\mathbf{z}_k, x^*)$ is independent of $S$ for each fixed $x^*$. Again, the actual prices charged to each insured will differ from those used to measure demographic parity.

\section{Discussion and Conclusion}
\label{sec:discussion}

We presented a unified framework for modeling transition rates in multi-state models using Poisson regression, and demonstrated how this framework enables the application of existing fair pricing methods to long-term insurance products. We illustrated the approach through a pricing exercise based on the HRS data, extending the post-processing methodology of \citet{lindholm2022discriminationfree} to the long-term insurance setting. We further discussed how the framework can be extended to the in-processing and pre-processing methodologies of \citet{beutel2017data} and \citet{lindholm2024proxy}, respectively.

While this paper focused on applying the unified Poisson-based framework to fair pricing in long-term insurance, the framework itself is substantially more general than the specifications used in our case study. We adopted log-linear models in our case study to reflect standard practice in multi-state modeling. However, a key contribution of the unified framework is that it removes the constraints imposed by the complexity of traditional multi-state likelihood functions, thereby accommodating more flexible functional forms for transition rates, such as that given by Equation~\eqref{eq:intensity2}. For example, an insurer employing our unified framework may choose to model transition rates as smooth functions of age using generalized additive models, aligning with the standard practice of smoothing and graduating age-specific mortality rates. This added flexibility opens the door to richer modeling choices beyond log-linear specifications,  offering a new direction for improving the modeling and pricing of complex products such as LTCI.

As regulators continue to assess and enforce fairness in insurance pricing, a key challenge is ensuring that chosen notions of fairness are applied coherently across different insurance products. One approach is to ensure fairness with respect to the price of each product. Thus, in adapting existing methodologies to the long-term insurance context, we have sought---wherever possible---to ensure that the final computed price also satisfies the enforced notion of fairness. As demonstrated in Section~\ref{sec:longtermfair}, doing so is possible for certain notions of fairness such as demographic parity, but requires careful formulation of the adapted methodology. It is unclear whether this ability to preserve the notion of fairness in the final price can be replicated for all existing notions or future proposed notions of fairness, but it is nevertheless a desired objective due to the coherence of defining fairness at the price level.  

\bigskip

\section*{Acknowledgments}

We acknowledge the financial support provided by the Society of Actuaries Research Institute.

\bibliographystyle{apalike}
\bibliography{reference}

@article{araizaiturria2024discriminationfree,
  title = {A discrimination-free premium under a causal framework},
  author = {Araiza Iturria, Carlos Andr{\'e}s and Hardy, Mary and Marriott, Paul},
  year = {2024},
  journal = {North American Actuarial Journal},
  volume = {28},
  number = {4},
  pages = {801--821},
  publisher = {Routledge},
  doi = {10.1080/10920277.2023.2291524},
  urldate = {2025-02-10}
}

@inproceedings{beutel2017data,
	title = {Data Decisions and Theoretical Implications when Adversarially Learning Fair Representations},
	author = {Alex Beutel and Jilin Chen and Zhe Zhao and Ed H. Chi},
	year = {2017},
	pages = {1-5},
series = {Fairness, Accountability and Transparency in Machine Learning '17},
numpages = {5},
location = {Halifax, Canada}
}

@book{barocas2023fairml,
  title = {Fairness and Machine Learning: Limitations and Opportunities},
  author = {Solon Barocas and Moritz Hardt and Arvind Narayanan},
  publisher = {MIT Press},
  year = {2023}
}

@article{lindholm2024proxy,
author = {Mathias Lindholm and Ronald Richman and Andreas Tsanakas and Mario V. W\"{u}thrich},
title = {What is fair? {{Proxy}} discrimination vs. demographic disparities in insurance pricing},
journal = {Scandinavian Actuarial Journal},
volume = {2024},
number = {9},
pages = {935--970},
year = {2024},
publisher = {Taylor \& Francis},
doi = {10.1080/03461238.2024.2364741},
URL = {https://doi.org/10.1080/03461238.2024.2364741}
}

@techreport{gabric2024bayesian,
  type = {Working {{Paper}}},
  title = {A {{Bayesian}} approach to discrimination-free insurance pricing},
  author = {Gabric, Lydia J. and Zhou, Shuang and Zhou, Kenneth Q.},
  year = {2024},
  institution = {Arizona State University},
  url = {https://papers.ssrn.com/abstract=4785927},
  urldate = {2024-07-25},
  langid = {english}
}

@techreport{grari2022fairml,
  type = {Working {{Paper}}},
  title = {A fair pricing model via adversarial learning},
  author = {Grari, Vincent and Charpentier, Arthur and Lamprier, Sylvain and Detyniecki, Marcin},
  year = {2022},
  url = {https://arxiv.org/pdf/2202.12008.pdf},
  urldate = {2024-07-24}
}

@techreport{hhs2020receiving,
  type = {Web page},
  title = {Receiving long-term care insurance benefits},
  author = {{Administration for Community Living}},
  year = {2020},
  address = {Washington, D.C.},
  institution = {{U.S. Department of Health and Human Services}},
  url = {https://acl.gov/ltc/costs-and-who-pays/what-is-long-term-care-insurance/receiving-long-term-care-insurance-benefits},
  urldate = {2021-08-27},
  langid = {english}
}

@article{hu2024seqfair,
  title = {A sequentially fair mechanism for multiple sensitive attributes},
  author = {Hu, Francois and Ratz, Philipp and Charpentier, Arthur},
  year = {2024},
  journal = {Proceedings of the AAAI Conference on Artificial Intelligence},
  volume = {38},
  number = {11},
  pages = {12502--12510},
  issn = {2374-3468},
  doi = {10.1609/aaai.v38i11.29143},
  urldate = {2024-07-25},
  copyright = {Copyright (c) 2024 Association for the Advancement of Artificial Intelligence},
  langid = {english}
}

@article{lindholm2022discriminationfree,
  title = {Discrimination-free insurance pricing},
  author = {Lindholm, M. and Richman, R. and Tsanakas, A. and W{\"u}thrich, M. V.},
  year = {2022},
  journal = {ASTIN Bulletin},
  volume = {52},
  number = {1},
  pages = {55--89},
  publisher = {Cambridge University Press},
  issn = {0515-0361, 1783-1350},
  doi = {10.1017/asb.2021.23},
  urldate = {2023-09-11},
  langid = {english}
}

@article{lindholm2024multitask,
  title = {A multi-task network approach for calculating discrimination-free insurance prices},
  author = {Lindholm, Mathias and Richman, Ronald and Tsanakas, Andreas and W{\"u}thrich, Mario V.},
  year = {2024},
  journal = {European Actuarial Journal},
  volume = {14},
  number = {2},
  pages = {329--369},
  issn = {2190-9741},
  doi = {10.1007/s13385-023-00367-z},
  urldate = {2024-07-23},
  langid = {english}
}

@techreport{colorado2022,
  type = {Senate {{Bill}} ({{SB}}) 21-169},
  title = {Restrict insurers' use of external consumer data},
  author = {{Colorado General Assembly}},
  year = 2021,
  address = {Denver, CO},
  url = {https://leg.colorado.gov/bills/sb21-169},
  urldate = {2025-12-29}
}

@techreport{nydfs2024,
  type = {Insurance {{Circular Letter No}}. 7},
  title = {Use of artificial intelligence systems and external consumer data and information sources in insurance underwriting and pricing},
  shorttitle = {Insurance circular letter no. 7 (2024)},
  author = {{New York State Department of Financial Services}},
  year = 2024,
  url = {https://www.dfs.ny.gov/industry-guidance/circular-letters/cl2024-07},
  urldate = {2025-12-29},
  langid = {english}
}

@inproceedings{dwork2012indfair,
author = {Dwork, Cynthia and Hardt, Moritz and Pitassi, Toniann and Reingold, Omer and Zemel, Richard},
title = {Fairness through awareness},
year = {2012},
isbn = {9781450311151},
publisher = {Association for Computing Machinery},
url = {https://doi.org/10.1145/2090236.2090255},
booktitle = {Proceedings of the 3rd Innovations in Theoretical Computer Science Conference},
pages = {214–226},
numpages = {13},
location = {Cambridge, Massachusetts},
series = {ITCS '12}
}

@article{cote2024fair,
author = {Côté, Olivier and Côté, Marie-Pier and Charpentier, Arthur},
title = {A fair price to pay: Exploiting causal graphs for fairness in insurance},
journal = {Journal of Risk and Insurance},
volume = {92},
number = {1},
pages = {33-75},
doi = {https://doi.org/10.1111/jori.12503},
url = {https://onlinelibrary.wiley.com/doi/abs/10.1111/jori.12503},
year = {2025}
}

@article{pessach2022review,
  title = {A review on fairness in machine learning},
  author = {Pessach, Dana and Shmueli, Erez},
  year = {2022},
  journal = {ACM Computing Surveys},
  volume = {55},
  number = {3},
  pages = {51:1--51:44},
  issn = {0360-0300},
  doi = {10.1145/3494672},
  urldate = {2024-07-25}
}

@article{wang2022multinn,
  title = {Multistate health transition modeling using neural networks},
  author = {Wang, Qiqi and Hanewald, Katja and Wang, Xiaojun},
  year = {2022},
  journal = {Journal of Risk and Insurance},
  volume = {89},
  number = {2},
  pages = {475--504},
  issn = {1539-6975},
  doi = {10.1111/jori.12364},
  urldate = {2023-09-03},
  copyright = {{\copyright} 2021 American Risk and Insurance Association},
  langid = {english}
}

@article{renshaw1995multipois,
title = {On the graduations associated with a multiple state model for permanent health insurance},
journal = {Insurance: Mathematics and Economics},
volume = {17},
number = {1},
pages = {1-17},
year = {1995},
issn = {0167-6687},
doi = {https://doi.org/10.1016/0167-6687(95)00011-G},
url = {https://www.sciencedirect.com/science/article/pii/016766879500011G},
author = {A.E. Renshaw and S. Haberman}
}

@article{sverdrup1965multipois,
author = {Erling Sverdrup},
title = {Estimates and test procedures in connection with stochastic models for deaths, recoveries and transfers between different states of health},
journal = {Scandinavian Actuarial Journal},
volume = {1965},
number = {3-4},
pages = {184--211},
year = {1965},
publisher = {Taylor \& Francis},
doi = {10.1080/03461238.1965.10405687}
}

@article{xin2024antidiscrimination,
  title = {Antidiscrimination insurance pricing: {{Regulations}}, fairness criteria, and models},
  shorttitle = {Antidiscrimination insurance pricing},
  author = {Xin, Xi and Huang, Fei},
  year = {2024},
  journal = {North American Actuarial Journal},
  volume = {28},
  number = {2},
  pages = {285--319},
  publisher = {Routledge},
  issn = {1092-0277},
  doi = {10.1080/10920277.2023.2190528},
  urldate = {2023-09-13}
}

@techreport{langa2020langaweir,
  type = {Survey {{Research Center}}},
  title = {Langa-{{Weir}} Classification of Cognitive  Function (1995 Onward)},
  author = {Langa, Kenneth M. and Weir, David R. and Kabeto, Mohammed and Sonnega, Amanda},
  year = {2020},
  address = {University of Michigan},
  institution = {Institute for Social Research},
  url = {https://hrsdata.isr.umich.edu/sites/default/files/documentation/data-descriptions/Data_Description_Langa_Weir_Classifications2016.pdf},
  urldate = {2025-08-03}
}

@article{fong2015disaggregating,
  title = {Disaggregating Activities of Daily Living Limitations for Predicting Nursing Home Admission},
  author = {Fong, Joelle H. and Mitchell, Olivia S. and Koh, Benedict S. K.},
  year = {2015},
  journal = {Health Services Research},
  volume = {50},
  number = {2},
  pages = {560--578},
  doi = {10.1111/1475-6773.12235},
  urldate = {2025-02-14},
  copyright = {{\copyright} Health Research and Educational Trust},
  langid = {english}
}

\appendix

\counterwithin{figure}{section}
\counterwithin{table}{section}


\section{Transforming transition data to support the unified framework}\label{sec:transformations}

We describe the procedure for transforming the transition data in Section~\ref{sec:hrsdata} to support the Poisson regressions required by the unified framework of Section~\ref{sec:unified}. 

\begin{table}[htbp]
\centering
\caption{Snippet of the transition dataset prior to any transformations.}
\label{tab:data1}
\begin{tabular}{l l r r r}
\toprule
Initial state & Ending state & Starting age & Ending age & Exposure\\
\midrule
Healthy & Disabled & 70.5 & 71.9 & 1.4 \\
Disabled & Dead & 71.9 & 73.8 & 1.9\\
\bottomrule
\end{tabular}
\end{table}

Consider two rows from the transition dataset prior to any transformations, as shown in Table~\ref{tab:data1}. Recall from our LTCI multi-state model that the type of transitions to which an insured is exposed depends on the insured’s initial health state: disability and healthy mortality (transitions 1 and 3, respectively) can occur only for healthy insureds, whereas recovery and disabled mortality (transitions 2 and 4, respectively) can occur only for disabled insureds. Our unified framework requires the creation of additional rows to indicate exposure to a given transition, and an event indicator to indicate whether the transition actually occurred. This transformation is illustrated in Table~\ref{tab:data2}. 

\begin{table}[htbp]
\centering
\caption{Snippet of the transition dataset after transformation to indicate exposure, corresponding to the data shown in Table~\ref{tab:data1}.}
\label{tab:data2}
\begin{tabular}{l r r r r}
\toprule
Transition & Starting age & Ending age & Event & Exposure\\
\midrule
1. Healthy $\to$ Disabled & 70.5 & 71.9 & 1 &  1.4 \\
3. Healthy $\to$ Dead & 70.5 & 71.9 & 0 &  1.4 \\
2. Disabled $\to$ Healthy & 71.9 & 73.8 & 0 & 1.9 \\
4. Disabled $\to$ Dead & 71.9 & 73.8 & 1 & 1.9\\
\bottomrule
\end{tabular}
\end{table}

The columns of transition type, event indicator, and exposure would be sufficient if transition rates were independent of age. However, we model transition rates as functions of age through the insured's age last birthday. Hence, coherent estimation of the transition rates using our proposed Poisson regression approach requires the age last birthday to remain constant within each row. We therefore create additional rows to split the observed period into time intervals over which this condition holds. For periods in which a transition occurs at the end of the observation window, the event indicator is assigned to the final interval, with all preceding intervals indicating no event, as illustrated in Table~\ref{tab:data3}. 

\begin{table}[htbp]
\centering
\caption{Snippet of the transition dataset after further transformation to ensure a constant age at last birthday for each row, following the transformation shown in Table~\ref{tab:data2}.}
\label{tab:data3}
\begin{tabular}{l r r r r r}
\toprule
Transition & Starting age & Ending age & Age &  Event & Exposure\\
\midrule
1. Healthy $\to$ Disabled & 70.5 & 71.0 & 70 & 0 &  0.5 \\
1. Healthy $\to$ Disabled & 71.0 & 71.9 & 71 & 1 &  0.9 \\
3. Healthy $\to$ Dead & 70.5 & 71.0 & 70 & 0 &  0.5 \\
3. Healthy $\to$ Dead & 71.0 & 71.9 & 71 & 0 &  0.9 \\
2. Disabled $\to$ Healthy & 71.9 & 72.0 & 71 & 0 & 0.1 \\
2. Disabled $\to$ Healthy & 72.0 & 73.0 & 72 & 0 & 1.0 \\
2. Disabled $\to$ Healthy & 73.0 & 73.8 & 73 & 0 & 0.8 \\
4. Disabled $\to$ Dead & 71.9 & 72.0 & 71 & 0 & 0.1 \\
4. Disabled $\to$ Dead & 72.0 & 73.0 & 72 & 0 & 1.0 \\
4. Disabled $\to$ Dead & 73.0 & 73.8 & 73 & 1 & 0.8 \\
\bottomrule
\end{tabular}
\end{table}

Once the transition dataset is restructured, the columns of starting age and ending age are no longer needed and can be discarded. The final step is to merge the covariate dataset with the restructured transition dataset. We first remove entries in the restructured dataset corresponding to individuals excluded during the covariate data cleaning process. The two datasets are merged using individual identifiers, with the relevant covariate information appended to each corresponding row.

The merged dataset is ready for the estimation of transition rates. Estimation for each transition type uses only the subset of observations corresponding to that transition; accordingly, the combined dataset is partitioned by transition type before being fed into the Poisson regression code.

\section{Model Selection for Illustrative GLM Model}
\label{app:simplifications}

To obtain a parsimonious GLM model for our case study in Section~\ref{sec:LTCIpricing}, we make the following changes:
\begin{itemize}
\item Marital status is consolidated into five categories: Married, Separated, Partnered, Widowed, Never. 

\item Years of education is grouped into the eight categories: did not complete elementary school (0-5 years), completed elementary but not middle school (6-7 years), completed middle school (8 years), some high school (9-11 years), completed high school (12 years), some college (13-15 years), completed college (16 years), and postgraduate (17+ years). 

\item Observations with census region coded as 5 are removed due to their scarcity (17 out of approximately 30,000 records), and because the U.S. Census Bureau defines only four census regions: Northeast, Midwest, South, and West.

\end{itemize}

In addition, we apply log-transformations to the financial variables to address their skewness. In particular, total household income is transformed using $\ln(1+x)$ before inclusion as a covariate. The treatment of total non-housing wealth is more complicated, as this variable can take negative values and therefore cannot be log-transformed directly. Letting $W$ denote the wealth, we find that the transformation $\ln(1+|W|)$ gives an approximately normal distribution. To capture both the magnitude and sign of wealth, we include three covariates: $\ln (1+|W|)$, $\sgn(W)$, and the interaction term $\sgn(W)\cdot \ln (1+|W|)$. 

\section{Likelihood Contributions of Fitted Covariates} 
\label{app:likelihood}

We measure the importance of each fitted covariate by the increase in likelihood it contributes to the model. Each contribution represents the additional likelihood gained by including that covariate after all other covariates have already been accounted for. The results are shown in Table~\ref{tab:likelihoods}. 

\begin{table}[htbp]
\centering
\begin{threeparttable}
\caption{Likelihood contributions for each of the covariates in the fitted GLMs.}
\label{tab:likelihoods}
\begin{tabular}{l rrrr r}
\toprule
 & \multicolumn{4}{c}{Transition} & \\
\cmidrule(lr){2-5}
Covariate & $H\to F$ & $F\to H$ & $H\to D$ & $F\to D$ & Total \\
\midrule
\multicolumn{6}{l}{\textit{Demographic factors}} \\
\addlinespace[0.5ex]
Age &2,360.23 & 1,587.60 & 2,031.35 & 2,498.03 & {8,477.21}\\
Years of education & 1,112.97 & 347.81 & 10.38 & 63.10 & {1,534.26}\\
Race & 695.94 & 40.99 & 1.79 & 21.12 & {759.84}\\
Labor force participation & 170.40 & 104.89 & 38.16 & 63.31 & 376.76\\
Gender & 115.30 & 0.18 & 140.09 & 68.62 & 324.19\\
Marital status  & 30.80 & 37.62 & 38.12 & 15.44 & 121.98\\
Ethnicity & 15.91 & 3.87 & 35.08  & 20.94 & 75.80\\
Census region & 41.74 & 3.03 & 11.87 & 2.69 & 59.33\\
Number of household residents & 0.22 & 2.31 & 1.91 & 0.29 & 4.73\\
\addlinespace[1ex]
\multicolumn{6}{l}{\textit{Health and disease factors}} \\
\addlinespace[0.5ex]
Self-rated health & 228.09 & 108.44 & 282.31 & 98.38 & {717.22}\\
Ever had diabetes & 23.13 & 27.14 & 95.10 & 159.14 & 304.51\\
Ever smoked cigarettes & 11.43 & 0.73 & 156.55 & 78.27 & 246.98\\
Ever had lung disease & 0.50 & 0.13 & 131.57 & 80.02 & 212.22\\
Ever drinks any alcohol & 29.83 & 28.94 & 10.49 & 10.57 & 147.57\\
Ever had stroke & 32.92 & 56.89 & 9.73 & 33.94 & 133.48\\
Ever had heart problems & 4.22 & 0.41 & 46.44 & 52.83 & 103.90\\
BMI & 0.26 & 8.28 & 13.68 & 33.29 & 55.51\\
\addlinespace[1ex]
\multicolumn{6}{l}{\textit{Financial variables}} \\
\addlinespace[0.5ex]
Non-housing wealth & 248.22 & 42.43 & 45.79 & 9.49 & 345.93 \\
Household income & 17.57 & 0.27 & 0.81 & 0.17  & 18.82\\
\bottomrule
\end{tabular}
\begin{tablenotes}[para]
\textit{Note}: $H$ stands for healthy, $F$ for (functionally) disabled, and $D$ for dead. Likelihood contributions are measured conditional on all other terms being included in the model. Variables are ordered by category, and in decreasing order of total likelihood contributions.
\end{tablenotes}
\end{threeparttable}
\end{table}

Across all categories of factors, the four most important factors---each with a total likelihood contribution exceeding 500---are age, years of education, race, and self-rated health, in decreasing order of importance. Age is by far the most important factor, with likelihood contributions far exceeding any other variable across all four transitions. This clearly establishes the primary role of age in driving transition rates. Years of education and race follow in importance, primarily influencing transitions through their effects on disability transition rates. Self-rated health ranks fourth, with a more evenly distributed likelihood contribution across all four transitions.

The importance of race is particularly notable given its role as the sensitive attribute of interest. The variable remains highly important despite controlling for various socioeconomic, health, and behavioral factors through the inclusion of other covariates. While race itself is unlikely to have any intrinsic effect on the transition rates, it may act as a proxy for other socioeconomic or health indicators not fully captured by the variables already included. 

\section{A Divide-and-Conquer Version of In-Processing}\label{app:madras}

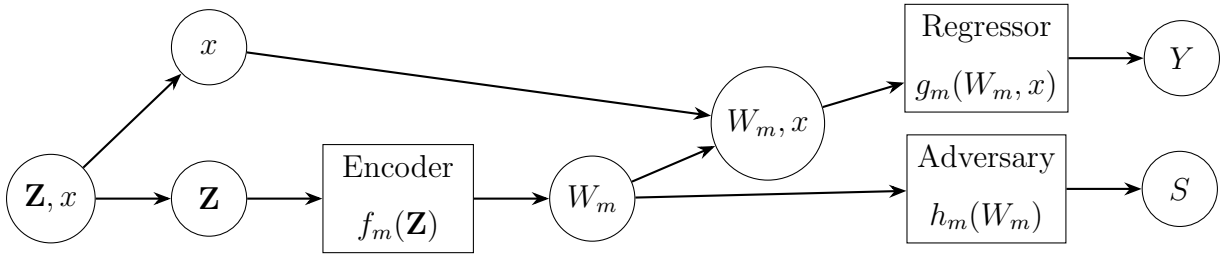
\begin{figure}[htbp]
\centering
\begin{tikzpicture}[
    node distance=1.5cm and 1cm,
    box/.style={draw, rectangle, minimum width=2cm, minimum height=1cm},
    circ/.style={draw, circle, minimum size=1cm},  
    arrow/.style={-Stealth, thick}
]

\node[circ] (Zx) {$\Z, x$};
\node[circ, right= of Zx] (Z) {$\Z$};
\node[circ, above= 1cm of Z] (x) {$x$};
\node[box, align = center, right= of Z](encoder){Encoder \\ $f_m(\mathbf{Z})$};
\node[circ, right=of encoder] (W) {$W_m$};
\node[circ, right=of W, yshift = 1cm] (Wx) {$W_m, x$};
\node[right = 2cm of Wx](mid){};
\node[box, align = center, above=0cm of mid] (regressor) {Regressor \\$g_m(W_m, x)$};
\node[circ, right=of regressor] (Y) {$Y$};
\node[box, align = center, below=0cm of mid] (adversary) {Adversary\\$h_m(W_m)$};
\node[circ, right=of adversary] (S) {$S$};
 \draw[arrow] (Zx) -- (x);
 \draw[arrow] (Zx) -- (Z);
 \draw[arrow] (Z) -- (encoder);
 \draw[arrow] (encoder) -- (W);
 \draw[arrow] (W) -- (Wx);
 \draw[arrow] (Wx) -- (regressor);
 \draw [arrow] (x) -- (Wx);
 \draw[arrow] (regressor) -- (Y);
 \draw[arrow] (W) -- (adversary);
 \draw[arrow] (adversary) -- (S);
\end{tikzpicture}
\caption{A version of the adversarial learning framework tailored to long-term insurance, adapted from \citet{beutel2017data} using the divide-and-conquer approach, shown for transition $m$. \textit{Note}: $\mathbf{Z}$ and $x$ refer to the non-age covariates and age, respectively; $W_m=f_m(\mathbf{Z})$ is the trained representation, and functions $f_m$, $g_m$ and $h_m$ are functions learned by the network.}
\label{fig:madras3}
\end{figure}

We describe a divide-and-conquer variant of the procedure presented in Section~\ref{sec:inprocessing}, with the corresponding architecture illustrated in Figure~\ref{fig:madras3}.

Repeat the following precedure for $m=1, 2, \ldots, M$: 
\begin{enumerate}
\item Begin with the sample of observed policies, with $\mathbf{z}_k$, $x_{u, k}$, and $s_k$ respectively denoting the vector of covariates, the age at entry, and the sensitive attribute associated with the $k$\textsuperscript{th} insured.
\item Let $w_{k, m} = f_m(\mathbf{z}_k)$. Using cross-entropy loss $L_S$, form the $m^\text{th}$ adversary loss
\begin{align*}
\mathit{Loss}_{Adv, m}&= \frac{1}{n}\sum_{k=1}^nL_S(s_k, h_m(w_{k, m})).
\end{align*}
\item Construct the dataset corresponding to transition $m$. Let $(k, m, j)$ index the $j$\textsuperscript{th} entry for the $k$\textsuperscript{th} insured and the $m$\textsuperscript{th} transition, with $n_{k, m}$ denoting the number of entries and $n_m = \sum_{k=1}^n n_{k, m}$. Using the Poisson likelihood loss $L_Y$, form the prediction loss for the $m$\textsuperscript{th} transition
\begin{align*}
\mathit{Loss}_{Pred, m} &= \frac{1}{n_m}\sum_{k=1}^n \sum_{j=1}^{n_{k, m}}L_Y(Y_{k, m, j}, g_m(w_{k, m}, x_{k, m, j})).
\end{align*}
\item Form the total loss for transition $m$
\begin{align}\label{eq:madrasloss3}
\mathit{Loss_m} &= \mathit{Loss}_{Pred, m} - \alpha \mathit{Loss}_{Adv, m}.
\end{align}
\item Alternate between gradient steps to {maximize} Equation~\eqref{eq:madrasloss3} with respect to $h_m$, and to minimize Equation~\eqref{eq:madrasloss3} with respect to $(f_m, g_m)$, until convergence.
\item Calculate the price for the $k$\textsuperscript{th} insured, $P(\mathbf{z}_k, {x_{u, k}})$, using  $\hat{\lambda}_m(\mathbf{z}, x) = g_m(f_m(\mathbf{z}), x)$. 
\end{enumerate}
The procedure ensures that demographic parity for the prices in the sense that $P(\mathbf{z}_k, x^*)$ is independent of $s_k$, for each fixed $x^*$. 



\end{document}